\documentclass[12pt]{iopart}
\usepackage{iopams}

\expandafter\let\csname equation*\endcsname\relax
\expandafter\let\csname endequation*\endcsname\relax
\usepackage{amssymb}
\usepackage{amsthm}
\usepackage{amstext}
\usepackage{mathtools}
\usepackage{graphicx}
\usepackage{placeins}
\usepackage{float}
\usepackage{subfigure}
\usepackage{CJK}
\usepackage{harvard}

\begin{document}
\begin{CJK*}{GBK}{ }

\title[Sia Nemat-Nasser]{Refraction Characteristics of Phononic Crystals}

\author{Sia Nemat-Nasser}

\address{Department of Mechanical and Aerospace Engineering\\
University of California, San Diego\\
La Jolla, CA, 92093-0416 USA}
\ead{sia@ucsd.edu}
\vspace{10pt}
\begin{indented}
\item[]November 25, 2014
\end{indented}
\end{CJK*}

\begin{abstract}

Some of the most interesting refraction properties of phononic crystals are revealed by examining the anti-plane shear waves in doubly periodic elastic composites with unit cells containing rectangular and/or elliptical multi-inclusions. The corresponding band-structure,  group velocity, and energy-flux vector
are calculated using a powerful mixed variational method which
accurately and efficiently yields all the field quantities over multiple frequency pass-bands.
The background matrix and the inclusions can be anisotropic, each having distinct elastic moduli and mass densities.
Equifrequency contours and energy-flux vectors are readily calculated as functions of the wave-vector components.
By superimposing the energy-flux vectors on equifrequency contours in the plane of the wave-vector components, and supplementing this with a three-dimensional graph of the corresponding frequency surface,
a wealth of information is extracted essentially at a glance.
This way it is shown that a composite with even a simple square unit cell containing a central circular inclusion can display negative or positive energy and phase-velocity refractions, or simply performs a harmonic vibration (standing wave),
depending on the frequency and the wave-vector.
Moreover that the same composite when interfaced with a suitable homogeneous solid can display:
\begin{enumerate}
\item negative refraction with negative phase-velocity refraction;
\item negative refraction with positive phase-velocity refraction;
\item positive refraction with negative phase-velocity refraction;
\item positive refraction with positive phase-velocity refraction; or even
\item complete reflection with no energy transmission,
\end{enumerate}
depending on the frequency, and direction and the wave length of the plane-wave which is incident from the homogeneous solid to the interface.

For elliptical and rectangular inclusion geometries, analytical expressions are given for the key calculation quantities. Expressions for displacement, velocity, linear momentum, strain and stress components, as well as the energy-flux and group-velocity components are given in series form.
The general results are illustrated for rectangular unit cells, one with two and the other with four inclusions, although any number of inclusions can be considered.
The energy-flux and the accompanying phase-velocity refractions at an interface with a homogeneous solid are demonstrated.

Finally, by comparing the results of the present solution method with those obtained using the Rayleigh quotient and, for the layered case, with the exact solutions, the remarkable accuracy and the convergence rate of the present solution method are demonstrated.

MatLab codes with comments will be provided.\\

\textit{Keywords}: Doubly periodic composites, phononic crystals, band structure, group and energy-flux vectors

\end{abstract}

\section{Introduction}
Periodic elastic composites exhibit phononic band structures and energy-flux patterns that depend directly on their micro-archtitectures and hence can be modified and controlled by micro-structural design.
The frequency band-structure in these composites results from the periodic modulation of field quantities, as in electronic band theory, and photonic and phononic crystals,
(\citeasnoun{bloch1928quantum},
\citeasnoun{brillouin1948wave},
\citeasnoun{rytov1956acoustical},
\citeasnoun{NN-1972a},
\citeasnoun{nemat1975harmonic},
\citeasnoun{minagawa1976harmonic}).

Such periodic modulations provide for very rich wave-physics and the potential for novel applications (\cite{cervera2001refractive},
\citeasnoun{yang2002ultrasound},
\citeasnoun{khelif2003trapping},
\citeasnoun{reed2003reversed},
\citeasnoun{yang2004focusing},
\citeasnoun{gorishnyy2005hypersonic},
\citeasnoun{mohammadi2008evidence},
\citeasnoun{sukhovich2008negative},
\citeasnoun{lin2009gradient}).
Equifrequency contours and energy-flux vectors are key information necessary to characterize the phononic response of periodic composites (phononic crystals), in exactly the same manner as for the photonic crystals (Ohtaka et. al, (1996) and Notomi (2000)).  Indeed, when the energy-flux vectors are displayed together with the equifrequency contours as functions of the wave-vector components, the direction of energy flow and the related phenomenon of positive or negative refraction are revealed essentially at a glance.
For this it would be necessary to efficiently and accurately calculate the composite's frequency pass-bands, displacement, velocity, momentum,  stresses, and group and energy-flux vectors as functions of the wave-vector components.

In the present paper we present a mixed variational formulation for phononic band-structure calculations, where both the displacement and the stress fields are varied independently and hence may be approximated by any  continuously differentiable  set of complete base functions, even though the displacement gradients may suffer large discontinuities across interfaces of various constituents of a typical unit cell.
Since the method  is based on a variational principle, any set of approximating functions can be used for calculations, e.g., plane-waves Fourier series or finite elements (\citeasnoun{minagawa1981finite}).
The method produces very accurate results and the rate of convergence of the corresponding series solution is greater than that of the  displacement-based approximating functions (\citeasnoun{babuska1978}).
Here we consider anti-plane shear waves and compare our results with those obtained using the Rayleigh quotient and clearly demonstrate both the accuracy and the speed of the convergence of our solution method. In addition, for a layered composite, we further show the  remarkable accuracy of our solutions.
Despite of its simplicity and effectiveness, the mixed-formulation has not yet been widely used to evaluate the band-structures of complex 2-, and 3-dimensional unit cells, even though such calculations existed in the literature since mid 1970's
(\citeasnoun{nemat1975harmonic},
\citeasnoun{minagawa1976harmonic}),
where composites with two-phase unit cells containing rectangular and/or ellipsoidal inclusions where analyzed and the corresponding band-structure and equifrequency contours were demonstrated.

\section{Statement of the Problem and Field Equations}

Consider a doubly periodic elastic composite composed of rectangular unit cells of common dimensions $a_1$ and $a_2$. A typical unit cell, $\Omega_1$, includes a nested set of concentric inclusions, $\Omega_l$, $l=2,3,...,n$, of rectangular or elliptical (or a combination of  both) shape,
$ \Omega_1\supset \Omega_2 \supset \Omega_3 ~  {...}  \supset \Omega_n.$
For simplicity let the principal axes of the inclusions be parallel to the coordinate axes,
${x}_1$ and ${x}_2$, although this is not necessary.

For Bloch-form time-harmonic anti-plane shear waves of frequency $\omega$ and wave-vector  components $k_1$ and $k_2$,
the dimensionless (see Section \ref{A1}) lateral displacement,
$w(\xi_1,\xi_2,t)$, and  in-plane shear stresses,
$\tau_j(\xi_1,\xi_2,t)$, $j=1,2$
have the following structure:
\begin{eqnarray}\label{fieldEQ1}
\left[ \begin{array}{c}
w\\
\tau_j\\
\end{array} \right]=
\left[ \begin{array}{c}
w^p(\xi_1,\xi_2)\\
\tau_j^p(\xi_1,\xi_2)\\
\end{array} \right]e^{i(Q_1\xi_1+Q_2\xi_2-\nu t)},
\end{eqnarray}
where $ \xi_j=x_j/a_j$, $Q_j=k_ja_j$ (no sum on $ j, ~ j =1,2$), $\nu$ is the dimensionless frequency (see Section \ref{A1}),
and superimposed $p$ denotes the periodic part.

The geometry and  the dimensionless mass-density, $\rho$, and the dimensionless elastic shear moduli,  $\mu_{jk}$
$j, k = 1, 2$, with $\mu_{12} = \mu_{21}$,
are periodic with the periodicity of the unit cell.
%
In what follows, $\mu_{12}=\mu_{21}=0$ is assumed. The normalized basic field equations now are,

\begin{eqnarray}
 	 \tau_{1,1}+a\tau_{2,2}+\nu^2\rho w=0; \label{field2}\\
	  w_{,1}=D_{11}\tau_1;\quad aw_{,2}=D_{22}\tau_2;\label{field3}\\
	  D_{11}=1/\mu_{11}\quad D_{22}=1/\mu_{22}; \label{field4}\\
      \tau_{1}=\mu_{11}w_{,1};\quad \tau_{2}=\mu_{22}aw_{,2};\label{field5}
\end{eqnarray}
where $D_{ij}$ is the normalized elastic compliance, and
comma followed by an index denotes differentiation with respect to the corresponding coordinate; the factor $a=a_1/a_2$ associated with $a(...)_{,2}\equiv a\frac{\partial(...)}{\partial{\xi_2}}$ is for non-dimensionalization of $x_j$ and the frequency, $\omega$, (see section \ref{A1}).


\section{Variational Formulation}

Consider now the following two functionals:
\begin{equation}
\lambda_N=\frac{<\tau_j,{w}_{,j}>+<{w}_{,j},\tau_j>-<D_{jk} \tau_k,\tau_j>}{ <\rho {w},{w}>};\label{NewQ}
\end{equation}
\begin{equation}
\lambda_R=\frac{<\mu_{jk}w_{,j},w_{,k}>}{<\rho {w},{w}>};\label{RayleighQ}
\end{equation}
where $<g u,v> = \int_{-1/2}^{1/2}\int_{-1/2}^{1/2}  guv^*d\xi_1d\xi_2$ for a real-valued function $g(\xi_1,\xi_2)$ and complex-valued functions $u(\xi_1,\xi_2)$ and $v(\xi_1,\xi_2)$, with star denoting complex conjugate, and
comma followed by the indexes 1 and 2 stand for
$\frac{\partial(...)}{\partial{\xi_1}}$ and
$a\frac{\partial(...)}{\partial{\xi_2}}$, respectively.
In (\ref{NewQ}) $w$ and $ {\tau}_j $ are viewed as independent fields subject to arbitrary variations, whereas in (\ref{RayleighQ}) $w$ is the independent field subject to arbitrary variation.
The first functional was introduced in early 1970's (\citeasnoun{nemat1972harmonic}) and was termed the \textit{new quotient}, and the second one is the well-known Rayleigh quotient.

It can be shown that the first variation of the new quotient for arbitrary variations of $w$ and $\tau_j$ with $\nu^2 = \lambda_N$ yields field equations (\ref{field2}, \ref{field3}) as the corresponding Euler equations, and that of the Rayleigh quotient for arbitrary variation of $w$  with $\nu^2 = \lambda_R$ yields
$(\mu_{jk}w_{,j})_k+\nu^2 w=0$, which is the displacement-based field equation obtained by substituting (\ref{field5}) into (\ref{field2}); note $\mu_{12}=\mu_{21} = 0$ is assumed.  We consider the new quotient first and then compare its results with those give by the Rayleigh quotient.

To find an approximate solution of the field equations (\ref{field2}, \ref{field3}) subject to the Bloch periodicity condition,
consider the following estimates:
\begin{equation}\label{Estimates}
{w}=\sum_{\alpha,\beta=-N}^{+N}W^{(\alpha\beta)}e^{i[(Q_1+2\pi\alpha)\xi_1+(Q_2+2\pi\beta)\xi_2]},
\end{equation}
\begin{equation}\label{Estimates2}
{\tau}_k=\sum_{\gamma,\delta=-N}^{+N}T_k^{(\gamma\delta)}e^{i[(Q_1+2\pi\gamma )\xi_1+(Q_2+2\pi\delta)\xi_2]},
\end{equation}
\begin{equation}
k=1,2,\quad
(k~not~summed).
\end{equation}
which automatically ensure the Bloch and continuity conditions.

Substitution into  (\ref{NewQ}) and minimization
with respect to the unknown coefficients $W$ and $T_j$ results in an eigenvalue problem which yields the band structure of the composite for anti-plane Bloch-form shear waves (see Section (\ref{A1}) for mathematical details),
\begin{equation}\label{Eigen-3}
\left[\Phi-\nu^2\Omega
\right]W=0,\qquad
det\left|\Phi-\nu^2\Omega\right|=0.
\end{equation}
Equation(\ref{Eigen-3})$_2$ yields the eigenvalues, $\nu$, as functions of  $Q_1$ and $Q_2$. Then, for each eigenvalue, $W$ is obtained from
equation (\ref{Eigen-3})$_1$; the corresponding  $T_j$ is then given by equation (\ref{T1}), Section (\ref{A1}).
\section{Group Velocity and Energy Flux}

The determinant in (\ref{Eigen-3})$_2$ depends parametrically on the wave-vector components, $Q_1\equiv k_1a_1$ and $Q_2\equiv k_2a_2$, where $a_1$ and $a_2$ are the unit cell dimensions.  The resulting eigenfrequencies, $\nu$, are thus functions of $Q_1$ and $Q_2$. These eigenfrequencies form surfaces in the ($Q_1$, $Q_2$,  $\nu$)-space, referred to as the Brillouin zones. The  fundamental zone corresponds to $-\pi{\leq{Q_1,Q_2}\leq }\pi$. We focus on this zone and examine the dynamic properties of the doubly periodic elastic composites on the first several frequency bands.

On each frequency band, the phase and group velocities are given by
\begin{equation}\label{ph-group}
v^p_{Jj}=\frac{\omega_Jk_j}{k_1^2+k_2^2},\quad
v^g_{Jj}=\frac{\partial \omega_J}{\partial{k_j}},\quad
j=1,2;
\end{equation}
here and below, $J=1, 2,...$ denotes the frequency band and $j=1,2$.
The refraction angle, say $\alpha_J$, is computed from
\begin{equation}\label{Refraction_Group}
\alpha_{J}=atan(\frac{v^g_{J2}}{v^g_{J1}}).
\end{equation}
It is known (\citeasnoun{brillouin1948wave}) that the direction, $\alpha_J$, is essentially the same as the direction of the energy flux for nondissipative media.  We shall illustrate this in what follows.

The $x_1$- and $x_2$-components of the (dimensionless) energy flux, averaged over a unit cell, are given by
\begin{equation} \label{Energy-flux}
\begin{split}
\bar{E}_{Jk} & =
\frac{\nu_J}{2\pi}
\int_{0}^{2\pi/\nu_J}
< Re(\tau_{kJ})
Re(\dot{w_J})^* > dt
=-\frac{1}{2}<\tau_{kJ}^p\dot{w}_J^{p*}>\\
& = \frac{1}{2}i\nu_J \sum_{\alpha\beta=-M}^{+M}T_{kJ}^{(\alpha \beta)}W_J^{(\alpha\beta)},\quad k=1,2,
\end{split}
\end{equation}
The direction, $ {\beta_J} $,  of the energy-flux vector is hence given by,
\begin{equation}\label{Refraction_Energy}
\beta_J=atan(\frac{\bar{E}_{J2}}{\bar{E}_{J1}}).
\end{equation}
It turns out that $\alpha_J=\beta_J$ for the class of problems considered in the present work.


In what follows, the above general results are illustrated in terms of several example.


\section{Two- and Four-phase Composites}\label{Exmp1}

We now examine the dynamic response of a two-phase phononic composite in some detail, and then briefly discuss the band structure of a four-phase  composite  to illustrate the results of the general formulation outlined above and detailed in Section (\ref{A1}).
In subsection (\ref{accuracy}) the convergence rate and the accuracy of the solutions are examined.

\begin{figure}[htp]
\centering
\includegraphics[scale=0.5, trim=0cm 0.0cm 0cm 0.5cm, clip=true]{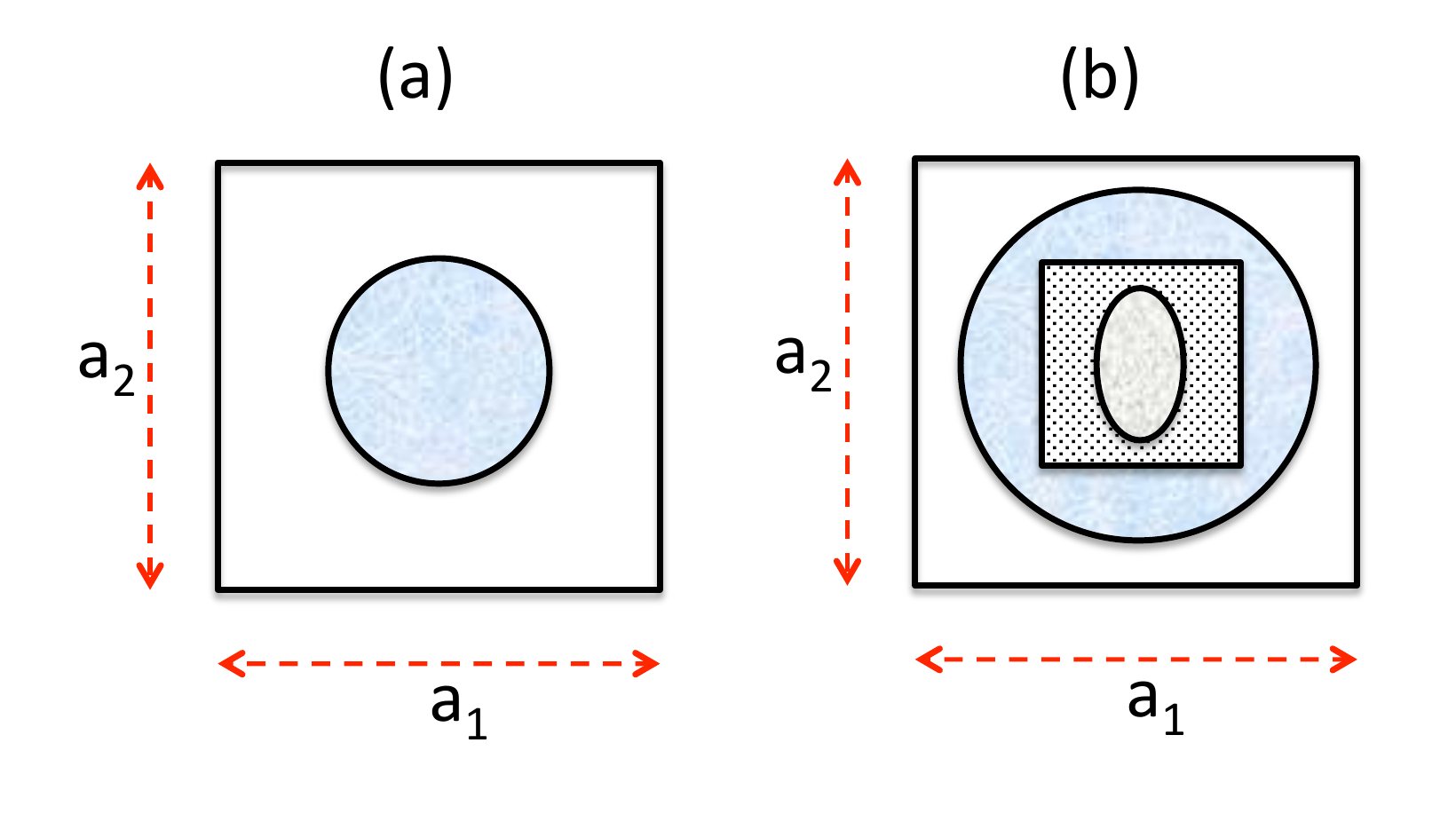}
\caption{ Unit cells: (a) a two-phase, and  (b) a three-phase composite.}
\label{Fig1new}
\end{figure}

The considered unit cells are shown in Figures  (\ref{Fig1new}a, b).
These examples show the rich body of physics that can be revealed by the present approach as well as  the versatility and effectiveness of the proposed computational tool.

\subsection{  Unit-cell Properties}\label{2ph-prop}

Consider the two-phase unit cell shown in Figure  (\ref{Fig1new}a).
The dimensionless parameters and the results are calculated using the following specific material properties (typical for PMMA and steel):
\begin{enumerate}
\item $\hat{\mu}_{1}$=1.7GPa; $\rho_{1}=1180$ kg/m$^3$; ${a}_1={a}_2$ = 5mm\label{polymer1}
\item $\hat{\mu}_{2}$=80GPa; $\rho_{2}=8000$ kg/m$^3$; $\hat{a}_1(2)=\hat{a}_2(2)$ = 1.5mm,\label{steel}
\end{enumerate}
where superimposed caret denote the dimensional quantity; see Section (\ref{A1}).


\subsubsection{Frequency Band Structure}\label{Bands1}

Now examine the variation of the frequency $f$ as a function of the normalized wave-vector components $Q_1$ and $Q_2$; for each pair of $Q_1$ and $Q_2$, the direction of the wave vector, the group-velocity vector, and the energy-flux vector are
obtained from
 $\theta=atan(aQ_2/Q_1)$, $\alpha_J=atan(v_{J2}^g/v_{J1}^g)$, and
 $\beta_J=atan(\bar{E}_{J2}/\bar{E}_{J1})$, respectively.
\begin{figure}[htp]
\centerline{\includegraphics[scale=.55, trim=0cm 0.5cm 0cm 0.1cm, clip=true]{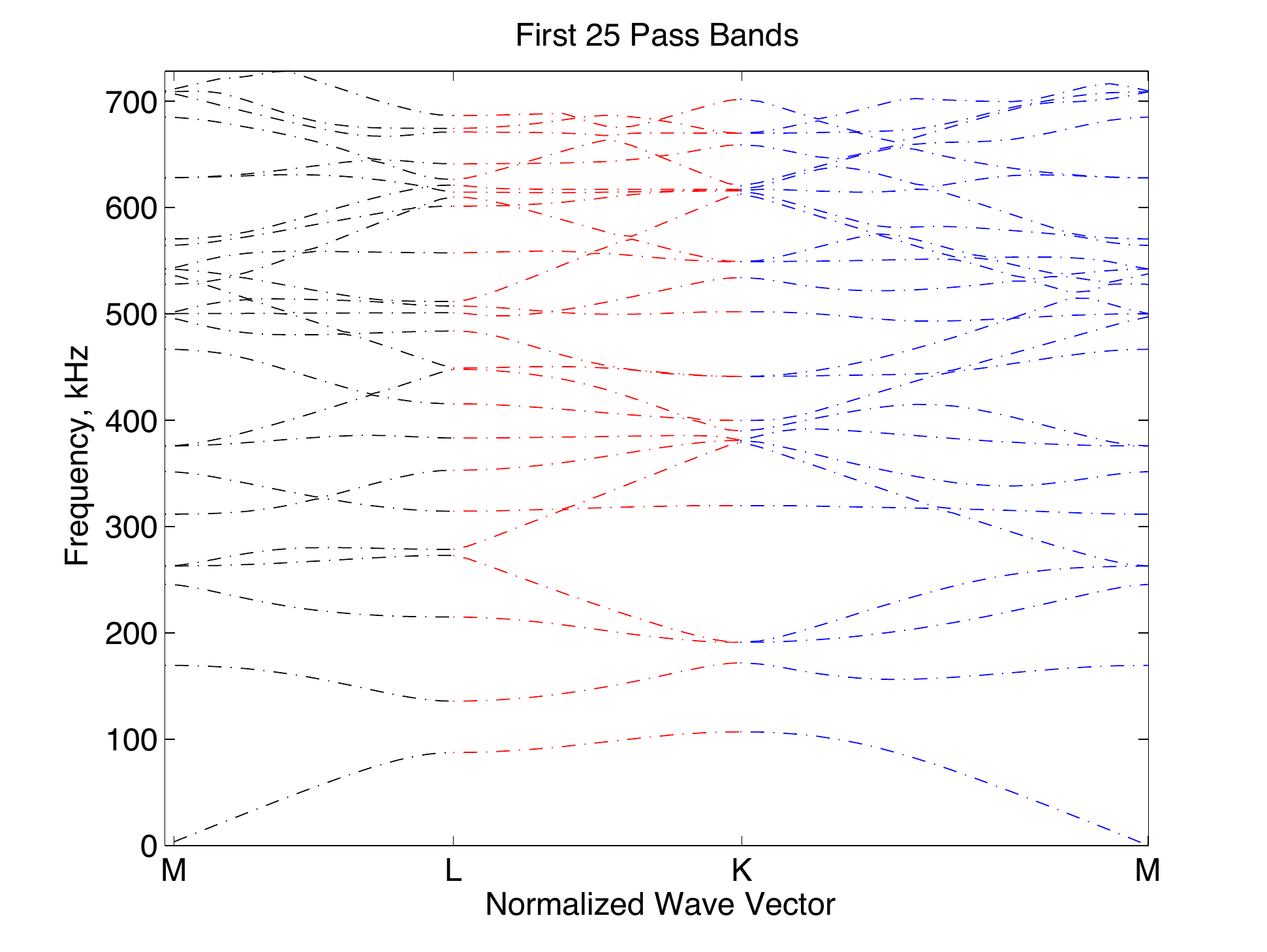}}
\caption{First 25 frequency pass-bands; two-phase unit cell with circular inclusion.}
\label{2p-25bandsMLKM}
\end{figure}

\begin{figure}[htp]
\centerline{\includegraphics[scale=.75, trim=0cm 2.5cm 0cm 2.1cm, clip=true]{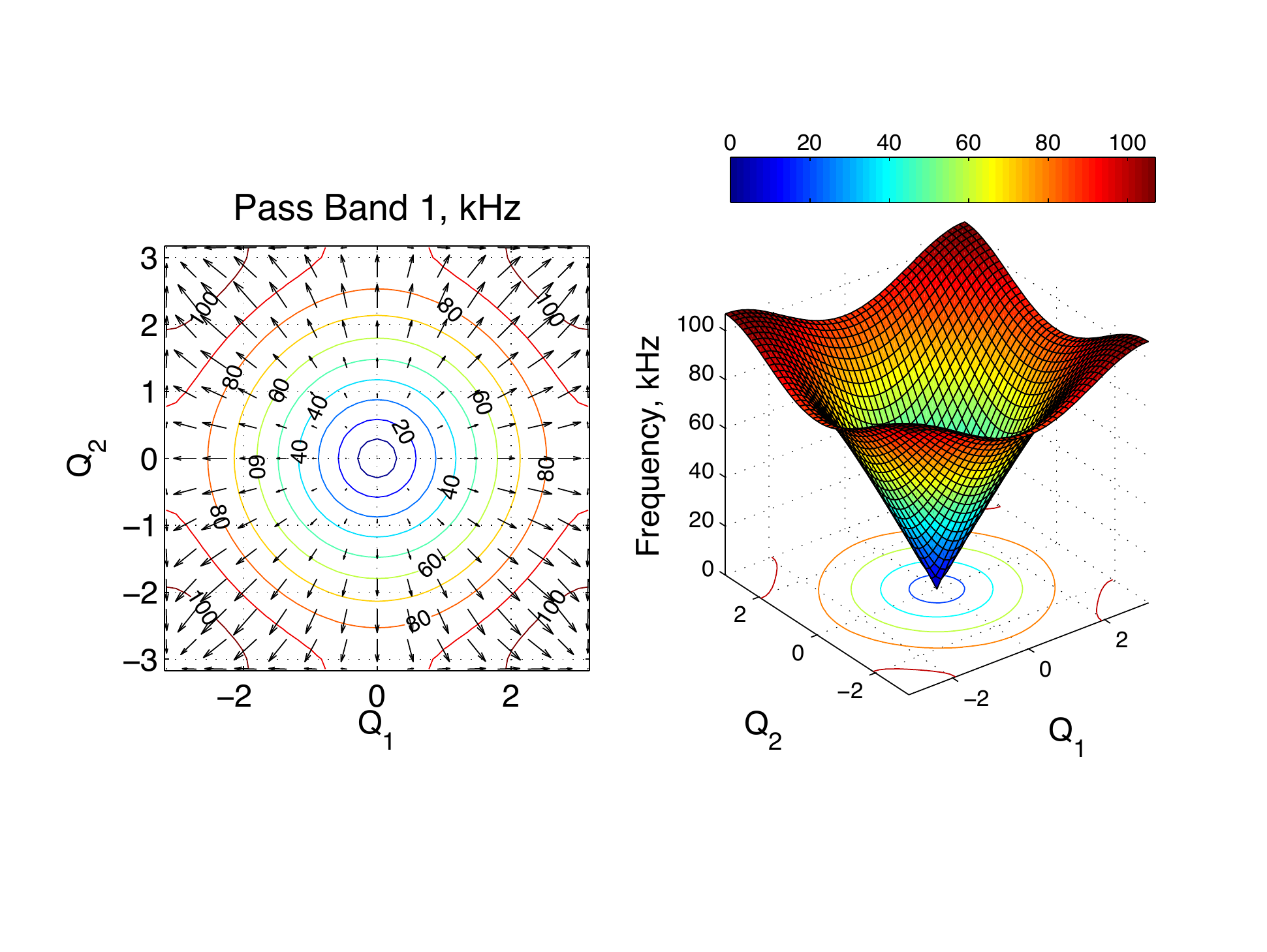}}
\caption{(Left) Equifrequency contours (in kHz) and energy-flux vectors, and
(Right) the corresponding three-dimensional graph with projected equifrequency contours;  first pass band.}
\label{1st-2p}
\end{figure}

\begin{figure}[htp]
\centerline{\includegraphics[scale=.75, trim=0cm 2.5cm 0cm 2.1cm, clip=true]{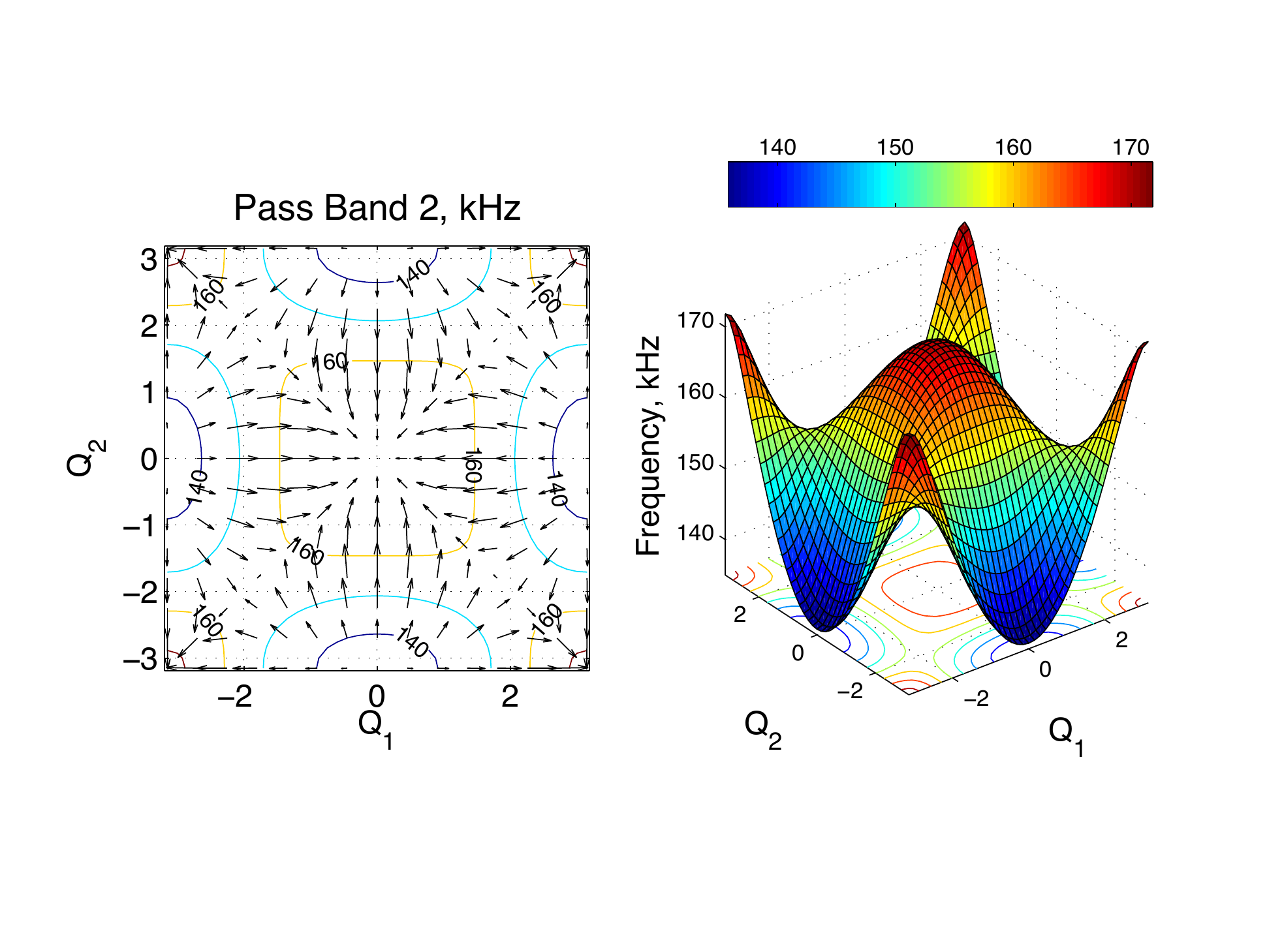}}
\caption{(Left) Equifrequency contours (in kHz) and energy-flux vectors, and
(Right) the corresponding three-dimensional graph with projected equifrequency contours;  second pass band.}
\label{2nd-2p}
\end{figure}

\begin{figure}[htp]
\centerline{\includegraphics[scale=.75, trim=0cm 2.5cm 0cm 2.1cm, clip=true]{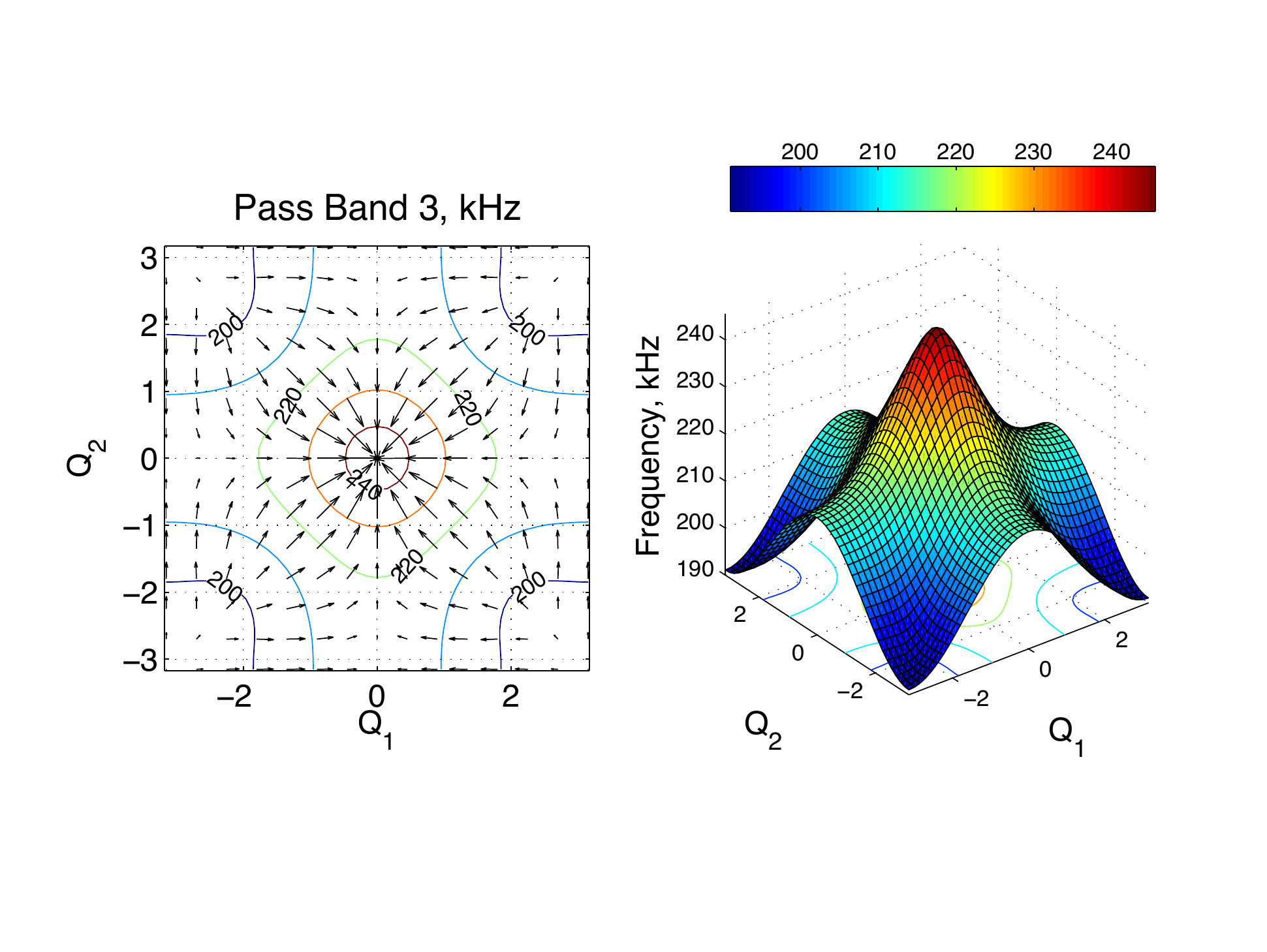}}
\caption{(Left) Equifrequency contours (in kHz) and energy-flux vectors, and
(Right) the corresponding three-dimensional graph with projected equifrequency contours;  third pass band.}
\label{3rd-2p}
\end{figure}

\begin{figure}[htp]
\centerline{\includegraphics[scale=.75, trim=0cm 2.5cm 0cm 1.95cm, clip=true]{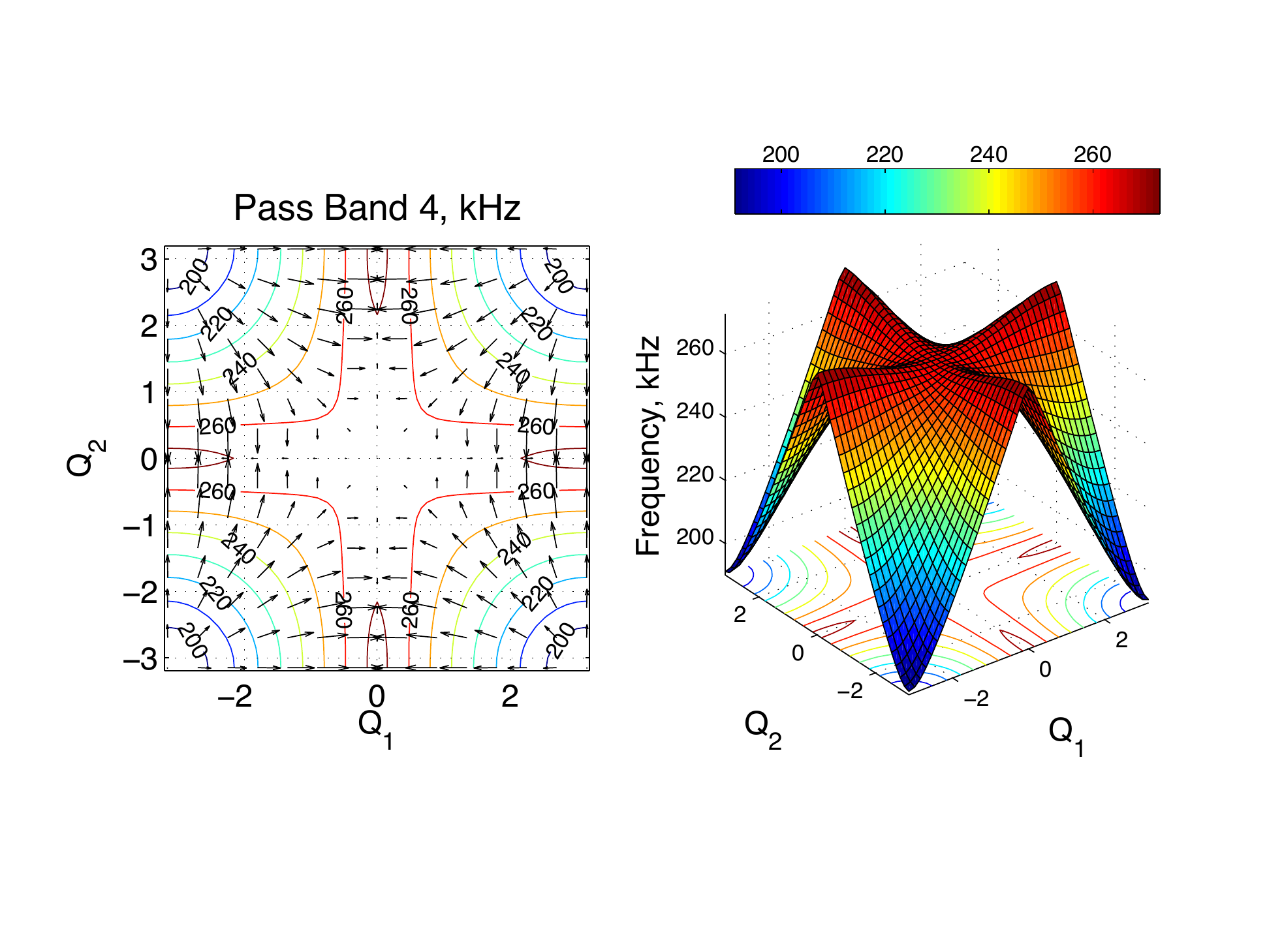}}
\caption{(Left) Equifrequency contours (in kHz) and energy-flux vectors, and
(Right) the corresponding three-dimensional graph with projected equifrequency contours;  fourth pass band.}
\label{4th-2p}
\end{figure}

The standard graph of the first 25 frequency pass-bands are given by
Figure (\ref{2p-25bandsMLKM}).
But the refraction characteristics of the composite is best revealed by the equifrequency contours in the $Q_1,Q_2$-plane with superimposed energy-flux vectors, accompanied by the corresponding three-dimensional frequency-graph, as are shown in
Figures (\ref{1st-2p}  to \ref{4th-2p}) for the first four frequency pass-bands.
As is seen, on the entire first pass band, Figure (\ref{1st-2p}),  energy flux vectors and the corresponding phase-velocity vectors are parallel, while they are antiparallel on the third band. Hence,
the composite would display positive refraction on its entire first pass-band and negative refraction on its entire third pass-band.
For points on the second pass-band, and to some extent on the fourth  pass-band, however, rich body of interesting refraction characteristics can be realized, as is discussed below.


\subsection{Wave-response on Second and Fourth Frequency Pass-bands}\label{secondB}

Consider first plane waves at $\pm \pi/4$ phase-velocity angle, i.e., $Q_2/Q_1=\pm 1$. For frequencies in the range of about 153 to 170 kHz
and $ |{Q_1}|=|{Q_2}|>1.93$, the phase-velocity and the energy-flux vectors are parallel: \textit{the composite does not display any refraction}. At around 153 kHz frequency and for phase-velocity vectors corresponding to $|Q_1|=|Q_2|\approx $1.93, the energy flux is zero.  Hence the composite could only support a \textit{standing wave}, i.e., perform free vibrations.
Then as the wave length increases with the corresponding frequency, and still for
$ |{Q_1}|=|{Q_2}|$, the energy-flux and the phase-velocity vectors become antiparallel. Therefore \textit{the composite would display negative refraction}.

When $Q_1=0$ and $Q_2\neq 0$ or $Q_2=0$ and $Q_1\neq 0$, the energy-flux and the wave-velocity vectors are antiparallel (\textit{negative refraction}). For other values of $Q_1$ and $Q_2$,  Figure (\ref{2nd-2p}) shows that the composite can display positive or  negative refraction depending on the associated frequency.  Many of the above-mentioned properties can be best understood by considering wave interaction at an interface of the composite with a suitable homogeneous solid, as is discussed in subsection (\ref{Bands2}) below.

Now examine the fourth frequency pass-band; Figure(\ref{4th-2p}).  As is seen, for frequencies and wave vectors associated with the points on the most part of this pass band, the composite would display negative refraction, with phase and energy-flux velocities being antiparallel, except for points on a  narrow region about $Q_1\approx \pm 0$ with $Q_2 \neq 0$ and $Q_2\approx \pm 0$ with $Q_1 \neq 0$ where the phase and group velocities are parallel (positive refraction).

\begin{figure}[htp]
\centerline{\includegraphics[scale=.65, trim=0cm 3.0cm 0cm 3.5cm, clip=true]{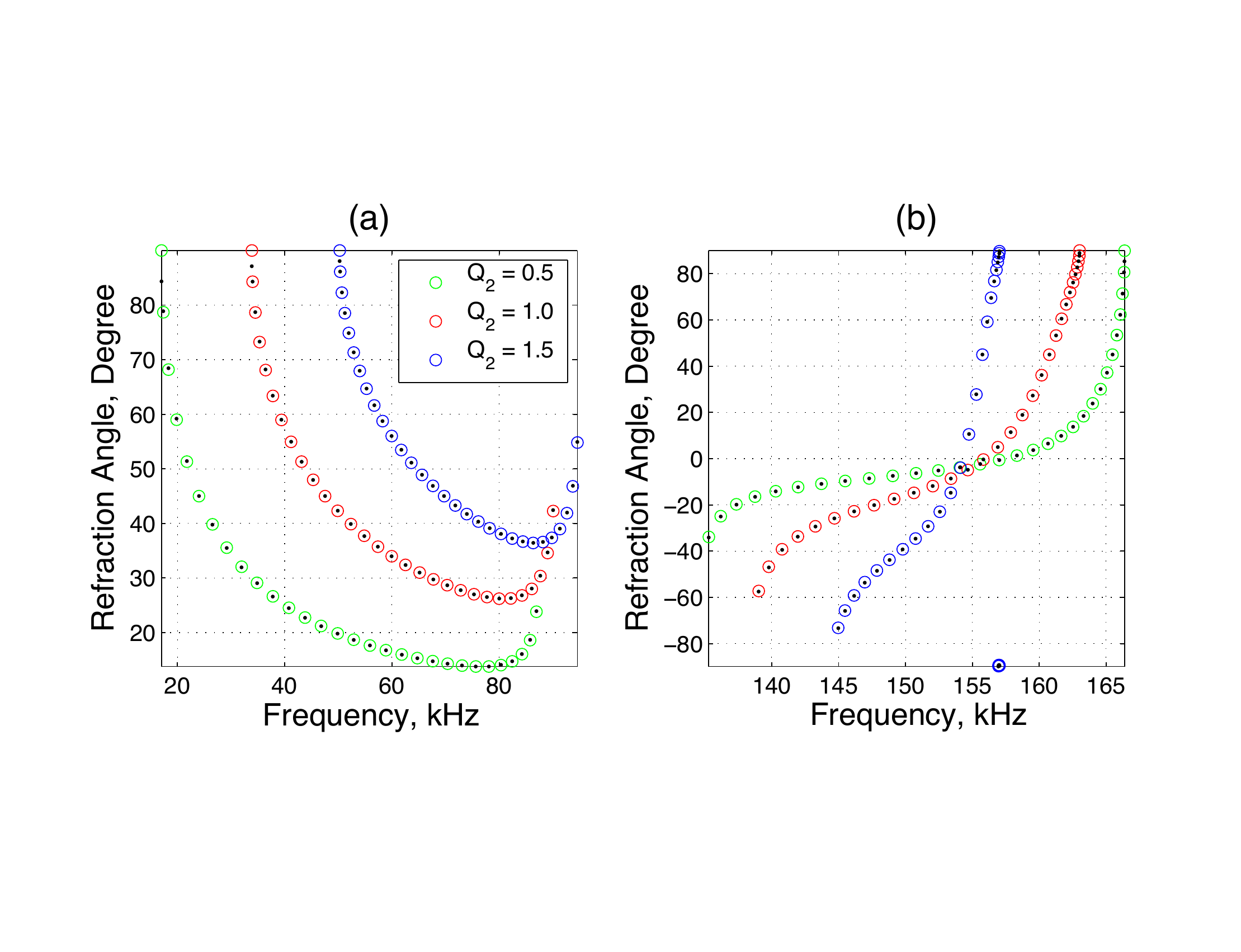}}
\caption{Energy-flux (green, red, and blue open circles ) and the corresponding group-velocity (black dots) directions for indicated values of $Q_2$: (a) first pass band, and (b) second pass band; two-phase unit cell with circular inclusion.}\label{2ph-EF_GV}
\end{figure}

As pointed out before, the direction of the energy-flux and group velocity vectors are essentially indistinguishable for this class of composites.
This is illustrated in Figure (\ref{2ph-EF_GV}) for $Q_2=0.5, 1.0, 1.5$, where the energy-flux directions are indicated by green, red, and blue open circles and the corresponding group-velocity directions by black dots, respectively.


\subsubsection{Refraction at an Interface: Second Pass-band}\label{Bands2}

\begin{figure}[htp]
\centerline{\includegraphics[scale=.5, trim=0cm 0.5cm 0cm 0.5cm, clip=true]{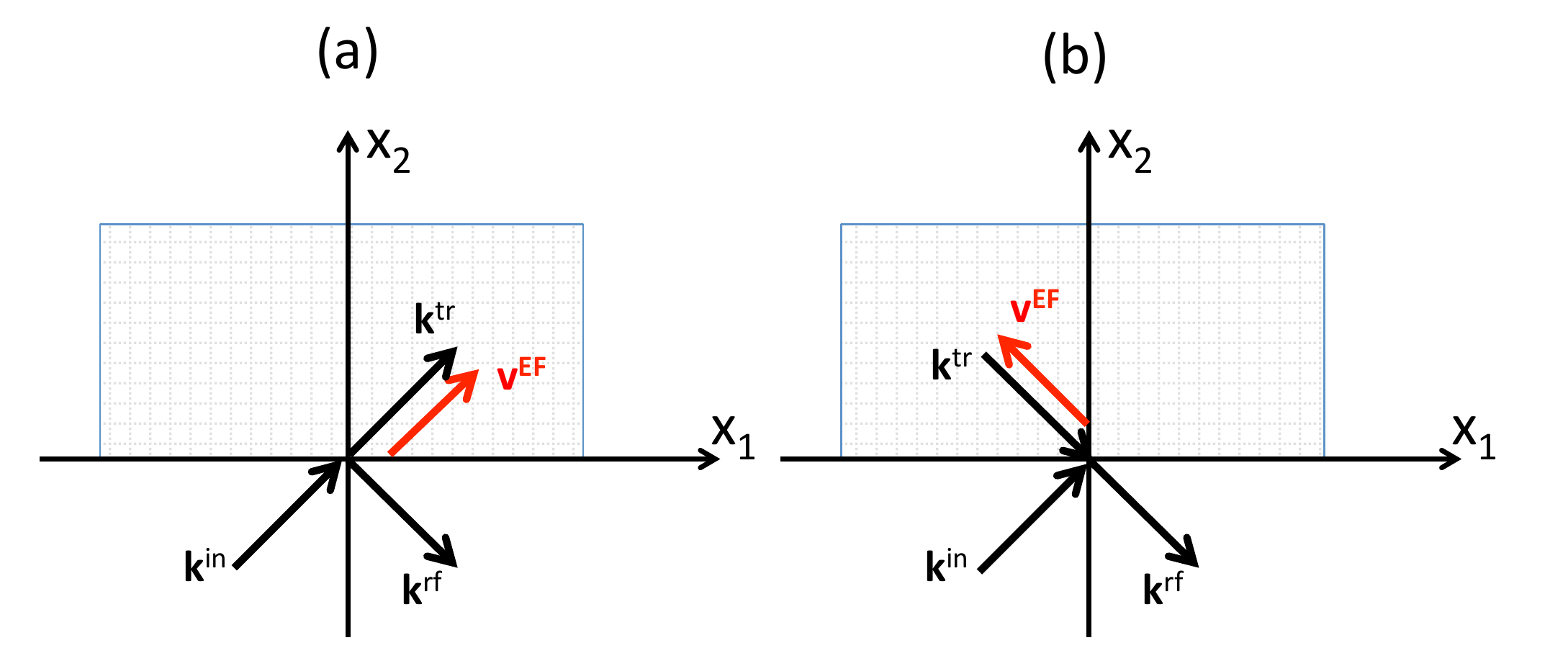}}
\caption{A plane wave is incident from a homogeneous half-space $x_2<0$ towards the interface $x_2 =0$ at a 45 degree angle. (a) For $Q_1^{in}>1.93$ and the incident wave frequency exceeding 153 kHz, transmitted energy-flux, $\mathbf{v^{EF}}$, and phase-velocity vectors are also at a 45 degree angle: \textit{no refraction}.  (b) For $Q_1^{in}<1.93$ and the incident wave frequency exceeding 153 kHz, $\mathbf{v^{EF}}$ and phase-velocity vector are at 135 degree and - 45 degree angles with respect to the $x_1$-axis, respectively: \textit{90 degree (negative) refraction}.}\label{interface-1}
\end{figure}

Let a plane wave be incident from a homogeneous half-space $x_2 <0$ towards the composite half-space, $x_2 >0$, at an angle of $\theta_0=atan (aQ_2^{in}/Q_1^{in})$, measured relative to the $x_1$-axis; in what follows, assume a square unit cell so that $a=a_1/a_2=1$.
Let $C_0$ be the shear-wave velocity of the
homogeneous solid and consider cases where $Q_1^{in}>0$.
At a (dimensionless) frequentcy of $\nu_0$, the homogeneous solid can support a shear wave  at an incident angle given
\begin{equation}\label{incidentAng}
cos(\theta_0)=
Q_1^{in}\frac{\bar{C}_0}{\nu_0},\qquad
\bar{C}_0=C_0\sqrt({\rho}/{\mu}_{11}).
\end{equation}
The continuity of the phase angle at the interface requires that
$Q_1^{tr} = Q_1^{in}$.
The corresponding $Q_2^{tr}$
is the root of the transcendental equation,
\begin{equation}\label{Q2tr}
\nu(Q_1^{in},Q_2^{tr})=\nu_0,
\end{equation}
which may be read off of the equifrequency contours in Figure (\ref{2nd-2p}).
If on the other hand, the value of $Q_2^{tr}$ is also fixed, the  value of the dimensionless frequency, $\nu_0$, would be given by equation (\ref{Q2tr}) and the corresponding incident angle by equation (\ref{incidentAng}), respectively.

For illustration examine the special case when
$Q_2^{tr}=Q_1^{tr}=Q_1^{in}$, say $=Q$. Then the frequency which could support this in the composite is given by $\nu_0=\nu(Q,Q)$ which corresponds to the incident angle
\begin{equation}\label{incidentAng0}
\theta_0= acos(\frac{Q\bar{C}_0}{\nu_0}).
\end{equation}
We now examine this case for various values of $Q$ and frequencies, $\nu_0$.
In Figure (\ref{2nd-2p}), these frequencies correspond to the
intersections of equifrequency
contours with either the upper or the lower diagonal line, for which the $x_2$-component of the energy-flux vector is positive; since the incident energy has to be taken away from the interface.
Depending on the value of $Q$, the following possibilities are identified:

(a) When $Q>1.93$, there would be positive refraction with positive phase refraction for incident-wave frequencies exceeding 153 kHz.
If in addition $\theta_0=45^o$, then there would be \textit{no refraction}:
the transmitted energy-flux vector,
$\mathbf{v^{EF}}$,
and the  transmitted wave-vector,
$\mathbf{k}^{tr}$, would be parallel, making a 45 degree angle with  the $x_1$-axis, the same as the incident wave; this is schematically shown in
Figures (\ref{interface-1}) .

(b) When $Q<1.93$, there would be negative refraction with positive phase refraction for incident-wave frequencies exceeding 153 kHz. In this case,
the transmitted energy-flux vector,
$\mathbf{v^{EF}}$, would be making a $90^o$ angle with
 the  transmitted wave-vector,
$\mathbf{k}^{tr}$, as suggested in Figures (\ref{interface-1}).
If in addition $\theta_0=45^o$, then there would be no phase refraction,
the transmitted phase-velocity vector being in line with the incident wave-vector, both making a $45^0$-angle with the $x_1$-axis, while the energy-flux vector is at a $135^o$ angle with respect to the $x_1$-axis (obliquely backward wave), as schematically shown in Figures (\ref{interface-1}).

(c) When $Q \approx 1.93$, there would be \textit{a total reflection with no energy transmitted} if the incident wave frequency is now about 153 kHz.
For this special case, the energy flux into the composite is zero. Hence the incident energy is taken away from the interface through total reflection.

Contrasting (c) above with the infinitely extended composite, note that, over a small window of specific frequencies and wave lengths, the \textit{infinitely extended composite can only perform free vibrations while a semi-infinite one behaves like a mirror at an interface with a homogeneous solid.}

\begin{figure}[htp]
\centerline{\includegraphics[scale=.5, trim=0cm 0.5cm 0cm 0.4cm, clip=true]{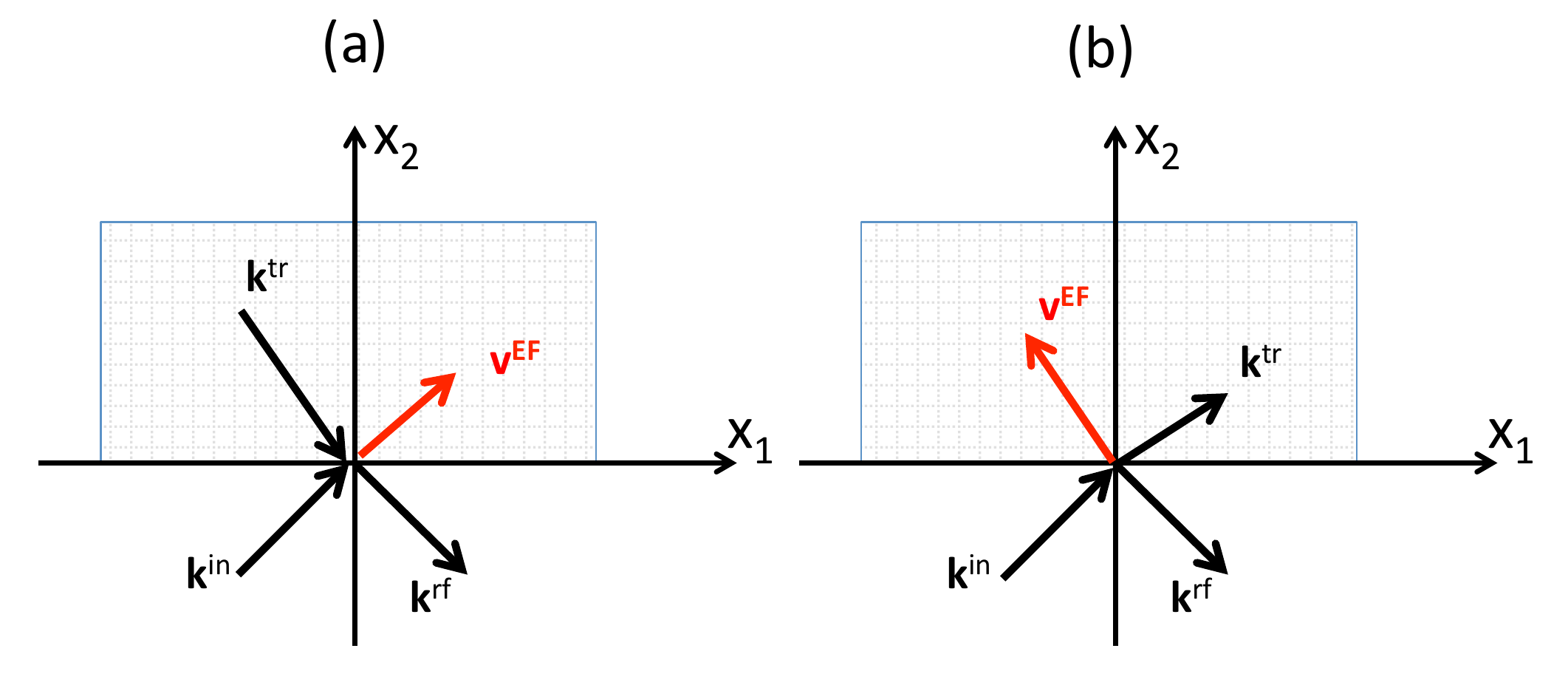}}
\caption{A plane wave is incident from a homogeneous half-space $x_2<0$ towards the interface
$x_2 =0$ at $\theta_0$ angle. (a) For $ Q_1^{in} = Q_1^{rf}=Q_1^{tr} = 1.65$ and the incident wave-frequency of about 151 kHz,  $Q_2^{tr} \approx$ - 2.35 and the transmitted energy-flux vector, $\mathbf{v^{EF}}$, and phase-velocity vector are  at about 41 degree and - 55 degree angles with respect to the $x_1$-axis: \textit{positive energy and negative phase refractions}.  (b) For $Q_1^{in}= Q_1^{rf}=Q_1^{tr}= $ 2.35, and the incident wave-frequency of about 151 kHz, $Q_2^{tr} \approx$ 1.65 and the  transmitted energy-flux vector, $\mathbf{v^{EF}}$, and phase-velocity vector are  at about 139 degree and 35 degree angles with respect to the $x_1$-axis: \textit{negative energy and positive phase refractions}.}\label{interface-2}
\end{figure}

Examine now two more rather interesting dynamic responses of this phononic crystal that Figure (\ref{2nd-2p}) readily reveals, and are illustrated in Figures  (\ref{interface-2}a, b).

In Figure (\ref{interface-2}a),
$Q_1^{in} = Q_1^{rf} = Q_1^{tr}$ = 1.65 and the incident frequency is 151 kHz. The composite supports this when $Q_2^{tr} $ = - 2.35, a negative phase-velocity at a - 55 degree angle. The corresponding energy-flux vector,
$\mathbf{v^{EF}}$, makes a 41 degree angle with the $x_1$-axis. Hence, the composite displays \textit{positive energy refraction with negative phase refraction}.

In Figure (\ref{interface-2}b),
$Q_1^{in} = Q_1^{rf} = Q_1^{tr}$ = 2.35 and the incident frequency is 151 kHz. The composite supports this when $Q_2^{tr} $ = 1.65, a positive phase-velocity at a 35 degree angle. The corresponding energy-flux vector,
$\mathbf{v^{EF}}$, makes a 139 degree angle with the $x_1$-axis. Hence, the composite displays \textit{negative energy refraction with positive phase refraction}.

Observe that negative refraction occurs when at least one component of the energy-flux  vector is antiparallel with one component of the corresponding phase-velocity vector. In
Figure (\ref{interface-1}b) both components are antiparallel, but in Figure (\ref{interface-2}b) only the $x_1$-components are antiparallel.  The first one is a case of \textit{complete backward wave} while the second is a case of \textit{obliquely backward wave}.

\subsection{Accuracy and Convergence Speed of Solutions}\label{accuracy}

\begin{figure}[htp]
\centerline{\includegraphics[scale=.65, trim=0cm 0.0cm 0cm 0cm, clip=true]{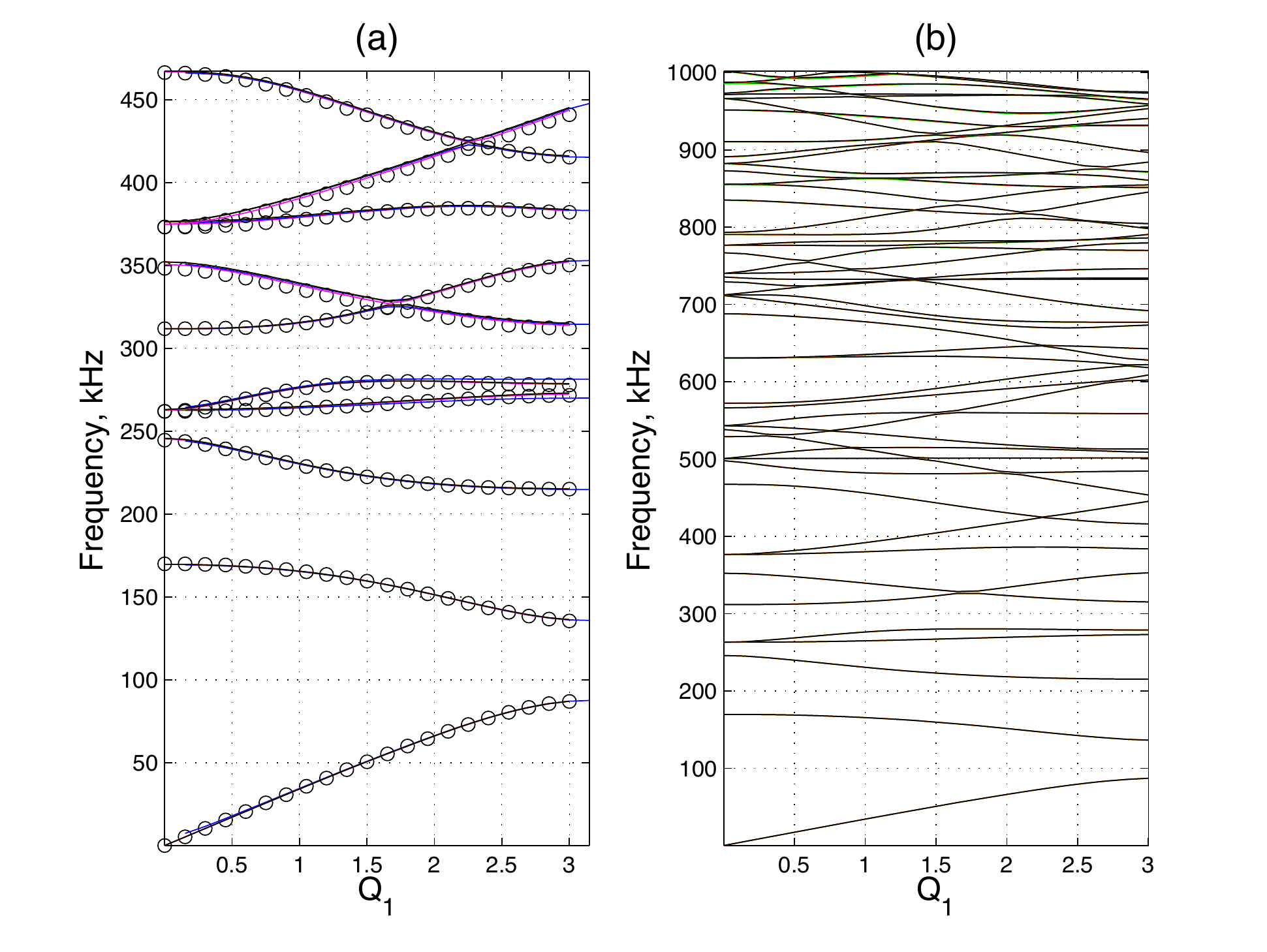}}
\caption{(a) First 10 frequency pass-bands for $N=$ 3 (open circles), and $N=$ 4, 5, 10, 12, 15 (magenta, blue, green, red, and black solid lines), respectively; and
(b) first 50 bands for $N=$ 10, 12, 15 (green, red, and black solid lines); $Q_2=0$.}\label{bands-2p}
\end{figure}

As a first step, examine the convergence rate of the series solution by comparing results corresponding to various values of the index $N$ which defines the number terms, $M=(2N+1)^2$, used in the series (\ref{Estimates}, \ref{Estimates2}).
In Figure (\ref{bands-2p}a) the first 10 bands are displayed for $N=3,4,5,10,12, 15$ and $Q_2=0$, using open black circles, and magenta, blue, green, red, and black solid lines, respectively. The solid lines are essentially indistinguishable. In Figure (\ref{bands-2p}b), the first 50 bands are displayed for $N=10,12, 15$ and the same value of $Q_2$, using the same corresponding colors.
We point out that for all relevant values of $Q_2$ the maximum difference here is less than 0.025\%.  Indeed even for the 100$^{th}$ band and relevant values of $Q_2$, the maximum difference is less than 0.45\%.
This however does not necessarily prove that the results have converged to the exact values.  This question is now addressed.

Rytov (1956) provided exact expressions for the Bolch-form harmonic waves in periodic elastic layers, a one-dimensional phononic crystal.  It is also know that the Rayleigh quotient, with the aid of a complete set of orthogonal base functions (Fourier series),  yields upper bounds for the eigenfrequencies (pass-bands) and that these bounds approach the exact results as the number of terms in the series solutions is increased.  Hence we can use both tools to examine the accuracy and the rate of convergence of our new quotient solution.

Consider first a layered phononic composite consisting of alternating (very compliant) polymer and (very stiff) steel sheets, having the following properties:
\begin{enumerate}
\item $\hat{\mu}_{1}$=0.5GPa; $\rho_{1}=700$ kg/m$^3$; ${a}_1= 2$mm
\item $\hat{\mu}_{2}$=80GPa; $\rho_{2}=8000$ kg/m$^3$; $\hat{a}_1= 3$mm.
\end{enumerate}

\begin{figure}[htp]
\centerline{\includegraphics[scale=.70, trim=0cm 0.0cm 0cm 1.0cm, clip=true]{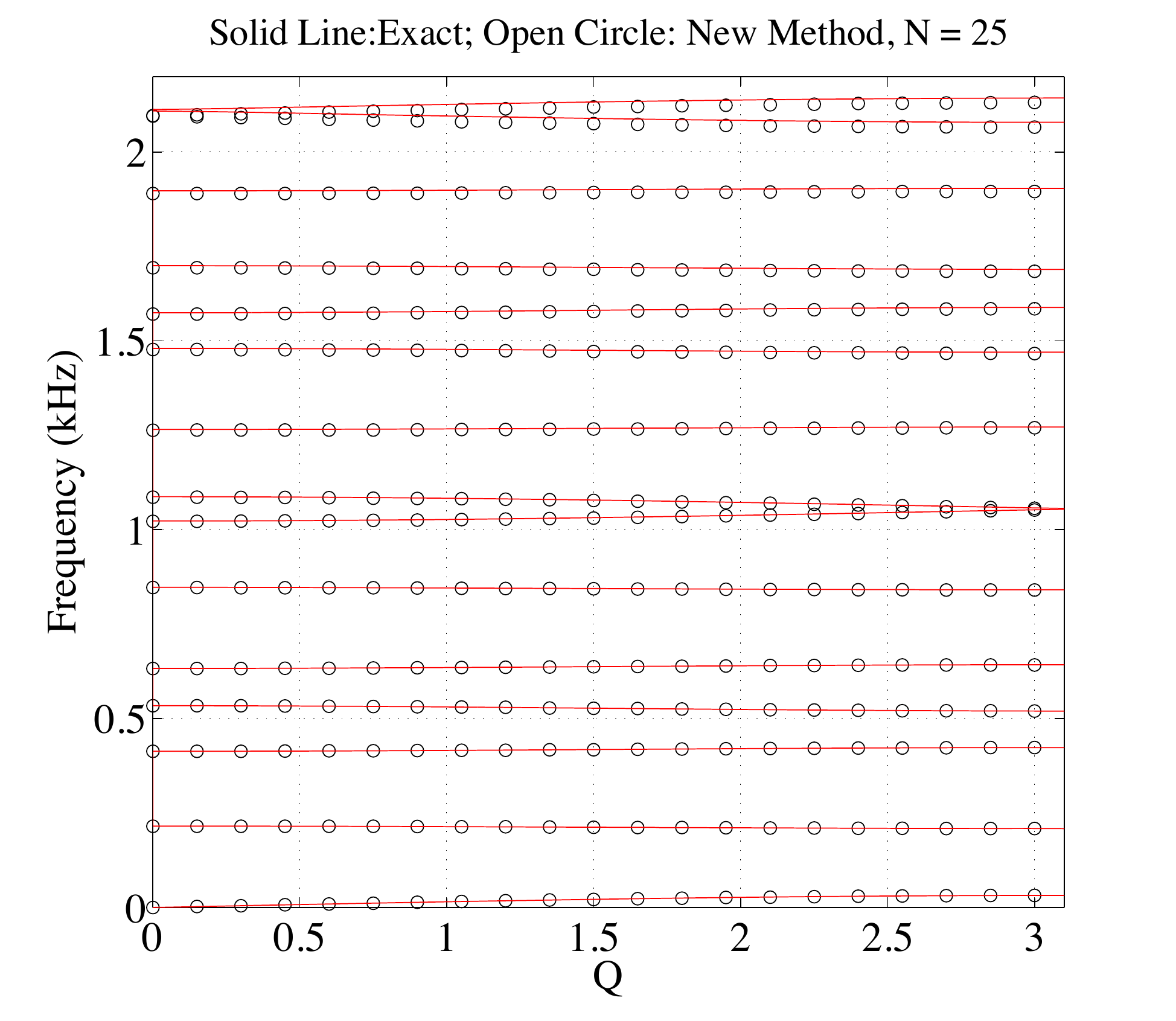}}
\caption{First 15 frequency pass-bands (in kHz) of a 2-phase layered  composite, comparing the exact  results (solid lines) with those of new quotient solution (open circles) for N = 25.}
\label{Rytov-NQ-1D-N25}
\end{figure}
The layers properties are chosen such that the ratio of the moduli to be rather large (here $ \hat{\mu}_{2}/\hat{\mu}_{1} =$160) to test the effectiveness of the new quotient solution method.
Figure (\ref{Rytov-NQ-1D-N25}) displays the first 15 pass bands for this layered composite. The solid lines are the exact results and the open circles are the results of the new quotient solution with $N=$ 25, i.e., $M=2N+1$ terms. As is seen the new quotient gives very accurate results. Note however that its results for higher bands tend to fall slightly below the exact values.  This is also the case for the two-dimensional phononic crystal discussed below.
\begin{figure}[htp]
\centerline{\includegraphics[scale=.70, trim=0cm 0.0cm 0cm .0cm, clip=true]{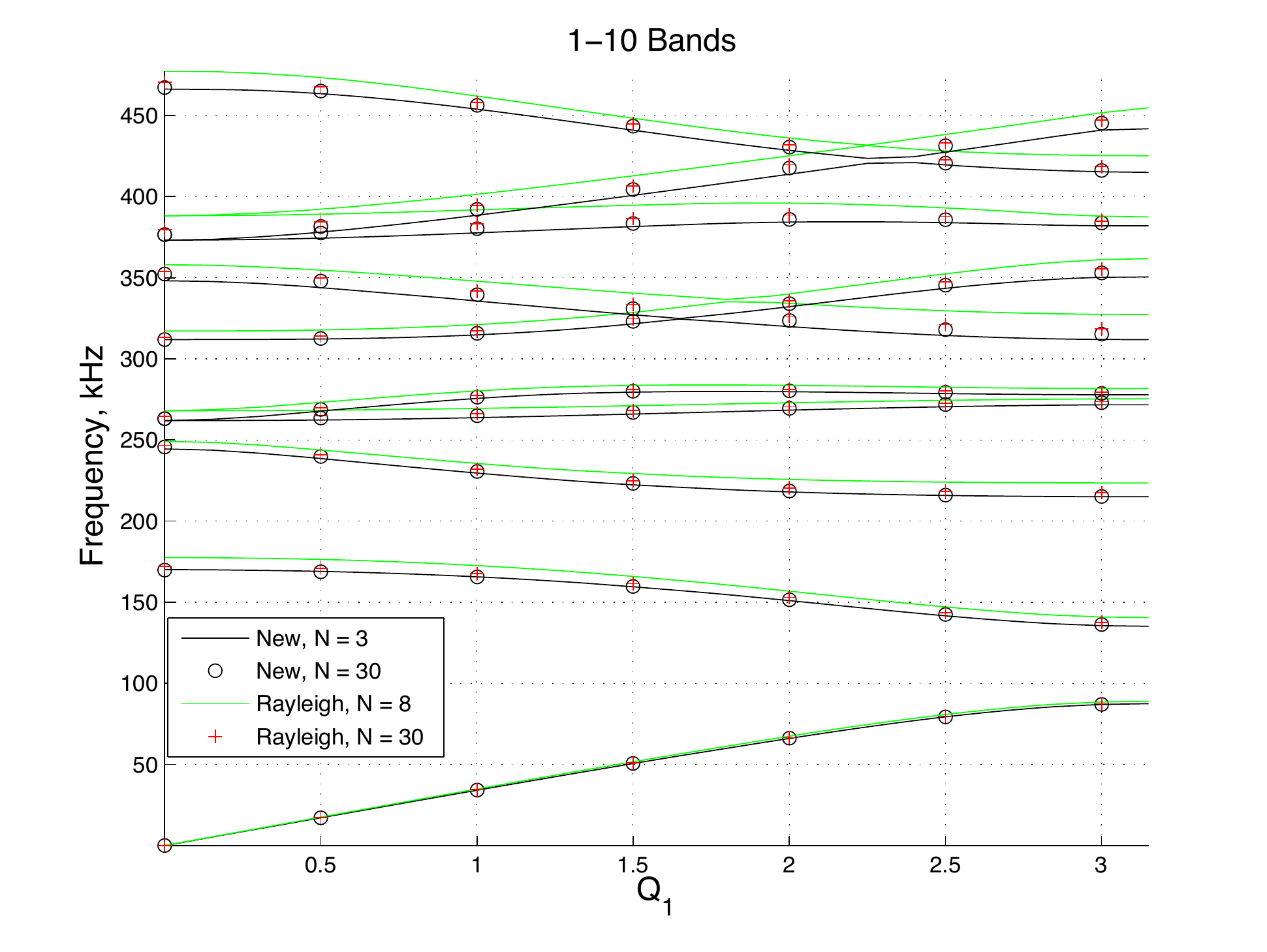}}
\caption{First 10 frequency pass-bands (in kHz) of the doubly periodic composite (Figure \ref{Fig1new}a ), comparing the results obtained by the Rayleigh quotient for N = 8, 30 (solid green lines and red crosses) with those obtained using the new quotient with N = 3, 30 (solid  black lines and open circles); $Q_2 =0$.}
\label{2D-10B-Rayleigh-New-for-paper}
\end{figure}

We now examine the accuracy of the solution given by our method for the  doubly periodic phononic crystal discussed in subsection (\ref{2ph-prop}).
Since there are no known exact solutions for this two dimensional case, we check our results against the upper bounds given by the Rayleigh quotient.

Figure (\ref{2D-10B-Rayleigh-New-for-paper})  displays the first 10 pass bands for $Q_2 = 0$, calculated using the Rayleigh quotient with $N=8, 30$ (solid green lines and red crosses) and the new quotient with $N=3, 30$ (solid black lines and open circles); there are, respectively, 49, 289, 3721 terms for $ N=3, 8, 30$ in the series solutions.
As is seen a 49-term series solution by the new quotient method yields estimates in par with the 3721-term series solution by the Rayleigh quotient, whereas the Rayleigh quotient gives poor results for even 289-term approximation; the figure also includes results of the new quotient for N = 30 for comparison.  Remarkably, the eigenvalue calculations for the new and the Rayleigh quotients require essentially the same computational efforts, as discussed in section (\ref{A1}).

\section{A Four-Phase Composite}\label{Exmp2}

To demonstrate the effectiveness of the computational platform, consider the four-phase unit cell shown in Figure  (\ref{Fig1new}b). The material properties and the cell dimensions are,
\begin{enumerate}
\item $\hat{\mu}_{1}$=0.5GPa; $\rho_{1}=700$ kg/m$^3$; $\hat{a}_1(1)=\hat{a}_2(1)$ = 8mm
\item $\hat{\mu}_{2}$=2.7GPa; $\rho_{2}=1180$ kg/m$^3$; $\hat{a}_1(2)=\hat{a}_2(2)$ = 6mm.
\item $\hat{\mu}_{3}$=0.4GPa; $\rho_{3}=700$ kg/m$^3$; $\hat{a}_1(3)=\hat{a}_2(3)$ = 3.5mm
\item $\hat{\mu}_{4}$=80GPa; $\rho_{4}=8000$ kg/m$^3$; $\hat{a}_1(4)=1.5$mm; $ \hat{a}_2(4)$ = 2.5mm
\end{enumerate}

\begin{figure}[htp]
\centerline{\includegraphics[scale=.75, trim=0cm 2.5cm 0cm 1.95cm, clip=true]{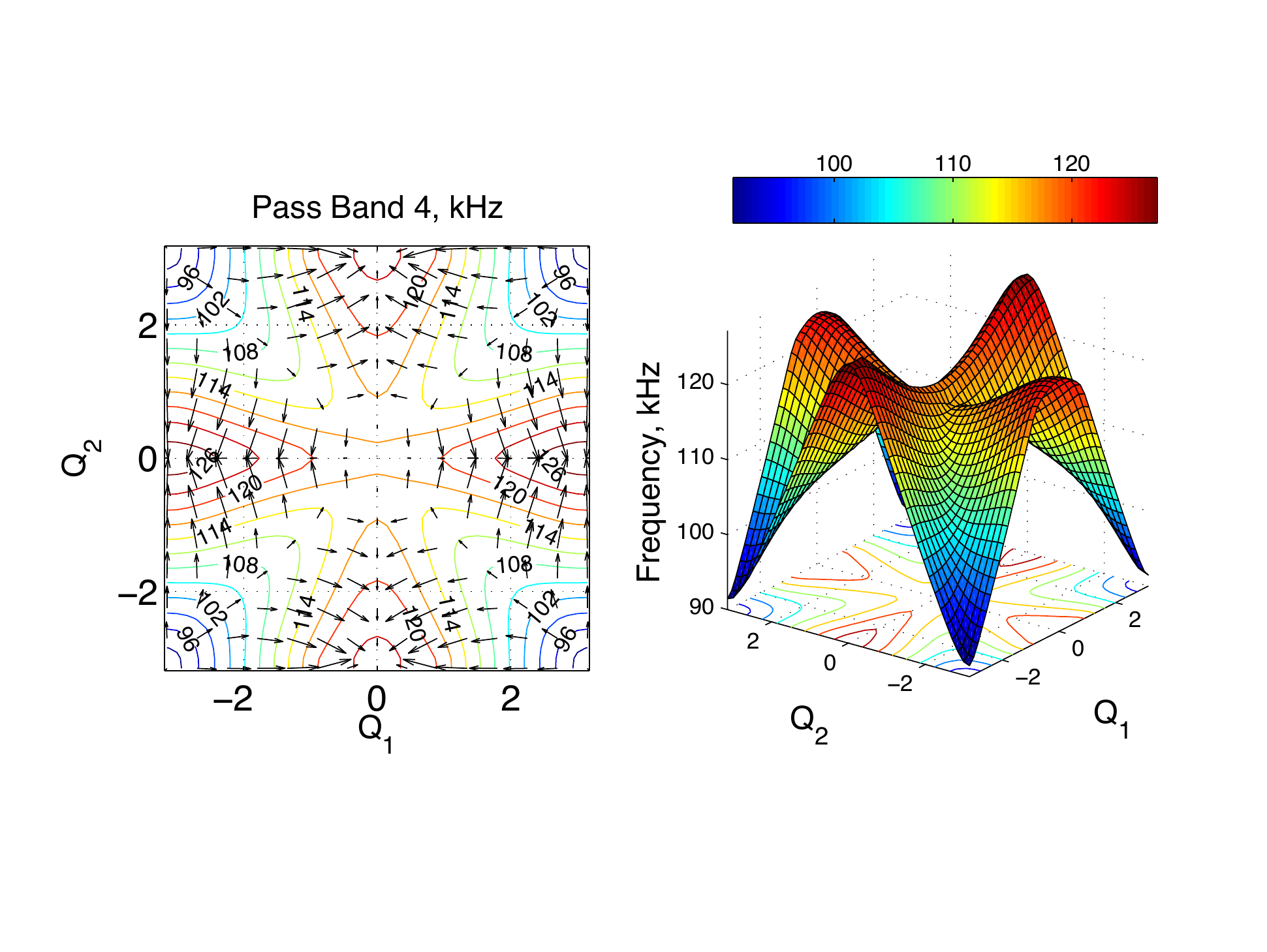}}
\caption{(Left) Equifrequency contours (in kHz) and energy-flux vectors, and
(Right) the corresponding three-dimensional graph with projected equifrequency contours;  fourth pass band.}
\label{4p-b4}
\end{figure}
\begin{figure}[htp]
\centerline{\includegraphics[scale=.75, trim=0cm 2.5cm 0cm 1.95cm, clip=true]{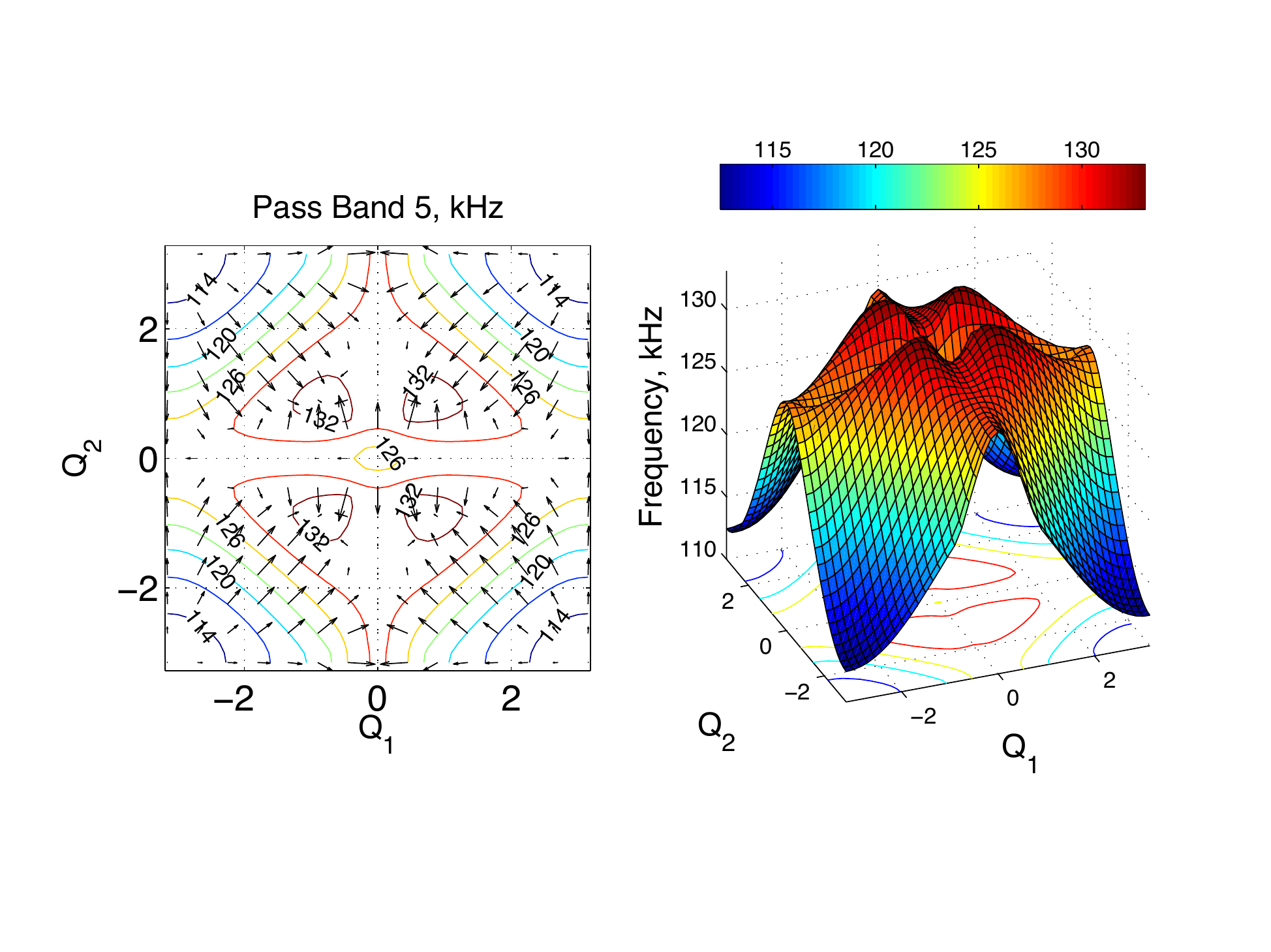}}
\caption{(Left) Equifrequency contours (in kHz) and energy-flux vectors, and
(Right) the corresponding three-dimensional graph with projected equifrequency contours;  fifth pass band.}
\label{4p-b5}
\end{figure}
\begin{figure}[htp]
\centerline{\includegraphics[scale=.75, trim=0cm 2.5cm 0cm 1.95cm, clip=true]{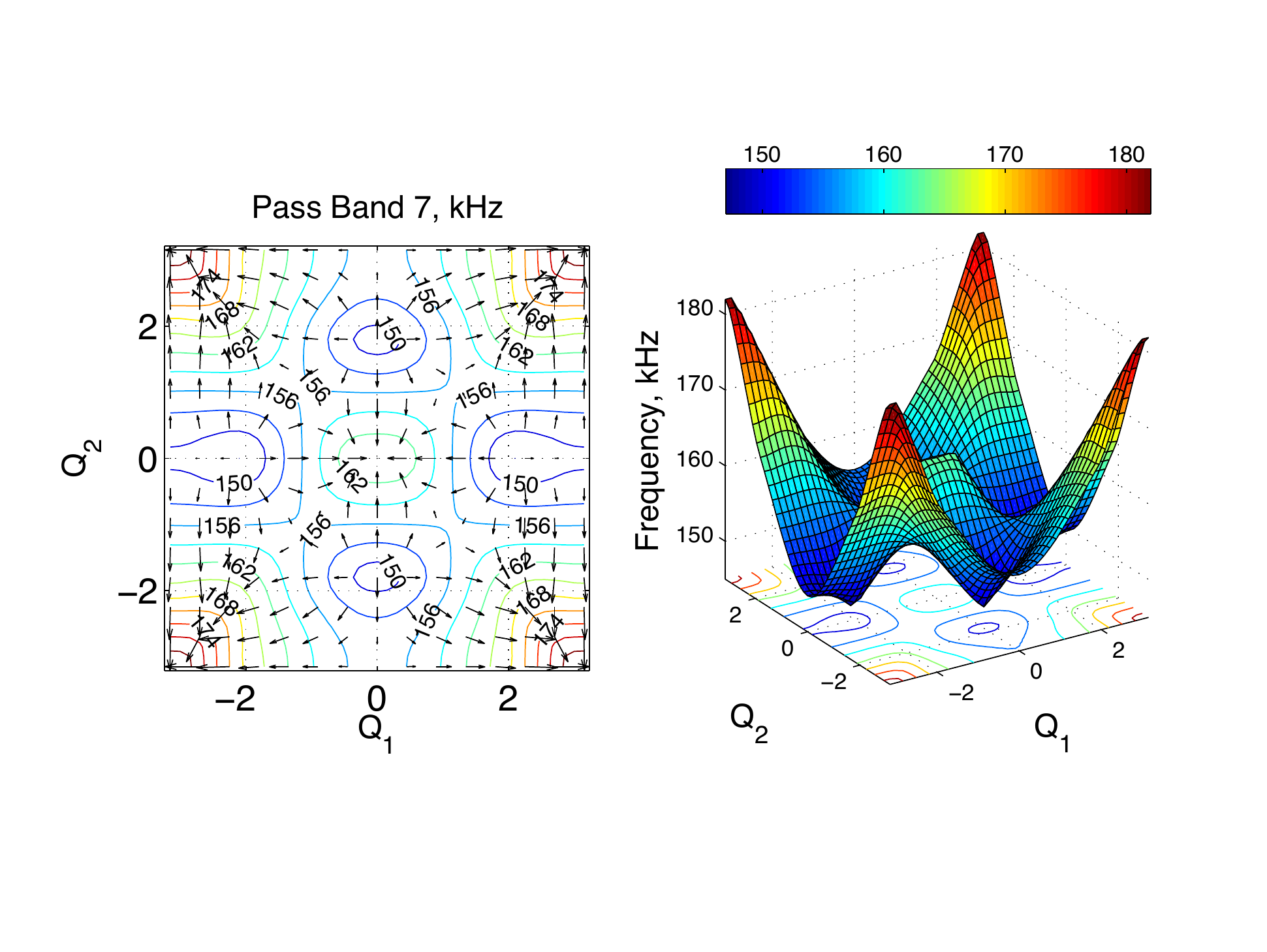}}
\caption{(Left) Equifrequency contours (in kHz) and energy-flux vectors, and
(Right) the corresponding three-dimensional graph with projected equifrequency contours;  seventh pass band.}
\label{4p-b7}
\end{figure}
The band structure and energy-flux pattern of the first four pass-bands are qualitatively similar to those of the two-phase composite considered above.
Here, the equifrequency contours are somewhat stretched in the $Q_1$-direction due to the anisotropy introduced by the elliptical shape of the central steel inclusion, as is seen in Figure (\ref{4p-b4}) which shows the fourth pass-band. The influence of the elliptical steel inclusion is more pronounce at  higher frequencies, as illustrated in Figures  (\ref{4p-b5}) and (\ref{4p-b7}). From these figures, one can immediately extract the entire refraction and other related wave-motion response of the composite when interfaced with a homogeneous solid.

\section{Calculation Details}\label{A1}
\subsection{Dimensionless Variables}

Let $\hat{a}_k(l)$, $k=1,2$ and $l=2,3,...,n$ denote the lengths of the principal axes of the inclusions, $\Omega_l$, with $\hat{a}_k(1)=a_k$.
Then  introduce the following dimensionless quantities,
\begin{equation}\label{Dimensionless-1}
a_1(l)=\hat{a}_1(l)/a_1,\quad
a_2(l)=\hat{a}_2(l)/a_2,\quad
l=2,3,...,n,\quad
a_1/a_2=a.
\end{equation}

Now, use some convenient reference elastic modulus $\tilde\mu$ and mass-density $\tilde\rho$ to normalize the variable density and elasticity of the unit cell, as well as the field quantities and the frequency, as follows:
\begin{equation}\label{DimensionLess-2}
\begin{split}
{\mu}_{jk}(\xi_1,\xi_2)=\hat{\mu}_{jk}/\tilde\mu,\quad
\rho(\xi_1,\xi_2)=\hat{\rho}/\tilde\rho,\quad
\tau_1=\sigma_{13}/\tilde\mu,\quad
\tau_2=\sigma_{23}/\tilde\mu,~~\\
\begin{aligned}
u=\hat{u}_3/a_1,\quad
\nu^2=a_1^2\omega^2\tilde\rho/\tilde\mu,\quad
Q_1=k_1a_1\quad
Q_2=k_2a_2,\quad
\xi_1=x_1/a_1,\\
\xi_2=x_2/a_2,\quad
u(\xi_1,\xi_2,t)= w^p e^{i(Q_1\xi_1+Q_2\xi_2-\nu t)},\quad
w=w^pe^{i(Q_1\xi_1+Q_2\xi_2)},
\end{aligned}
\end{split}
\end{equation}
where $\nu$ is the dimensionless frequency.  Here the displacement, $\hat{u}_3$, and the nonzero (shear) stresses, $\sigma_{13}=\sigma_{31}$ and $\sigma_{23}=\sigma_{32}$, are rendered nondimensional and denoted by $w$ and $\tau_j$, $j=1,2$, respectively, where $w^p$ is the periodic part of the displacement field. Also, the strains, $2\epsilon_{13}=2\epsilon_{31}$ and $2\epsilon_{23}=2\epsilon_{32} $,  will henceforth be denoted by $\gamma_1$ and $\gamma_2$, respectively.

\subsection{Frequency-band Calculations}\label{B1}

We first discuss the new quotient and then show in subsection (\ref{Rayleigh}) how by a simple modification one can obtain the results for the Rayleigh quotient.

Direct substitution of (\ref{Estimates}, \ref{Estimates2}) into (\ref{NewQ}) and minimization with respect to $W$ and $T_j$ as the independent variables yields,
\begin{eqnarray}\label{Euler1}
\left[ \begin{array}{ccc}
iH_1 & iH_2 & \nu^2 \Lambda_{\rho} \\
\Lambda_{D_{11}} & 0 & -iH_1\\
0 & \Lambda_{D_{22}} & -iH_2
\end{array} \right]\
\left[ \begin{array}{c}
T_1\\
T_2\\
W\end{array} \right]=0,
\end{eqnarray}
where $\Lambda_f=[\Lambda^{(\alpha \beta,\gamma  \delta)}_{f}]$ is an
$M^2 \times M^2$ matrix, and
$H_{1}$ and $H_{2}$ are two $M^2\times M^2$ diagonal matrices with the respective components
$(Q_1+2\pi \alpha)\delta_{\alpha \gamma}$ and
$(Q_2+2\pi \alpha)a\delta_{\beta \delta}$.
The components of $\Lambda_f$ are defined by
\begin{equation}\label{Operator}
\Lambda^{(\alpha \beta,\gamma  \delta)}_{f}=
 \int_{-1/2}^{1/2}\int_{-1/2}^{1/2} f(\xi_1,\xi_2)
e^{i2\pi [(\alpha - \gamma)\xi_1+(\beta -\delta)\xi_2]}d\xi_1 d\xi_2,
\end{equation}
with $f(\xi_1,\xi_2)$ being a real-valued integrable function.
For an even function, $f(\xi_1,\xi_2)=f(-\xi_1,-\xi_2)$ (symmetric unit cells),
and
$\Lambda^{(\alpha \beta,\gamma  \delta)}_{f}=
\Lambda^{(\gamma  \delta,\alpha \beta)}_{f}$ is real-valued.

From the system of linear and homogeneous equations (\ref{Euler1}),we obtain,
\begin{equation}\label{Eigen1}
\left[
\Phi-\nu^2\Omega
\right]W=0,\quad
\Phi=(H_1 \Lambda_{D_{11}}^{-1} H_1+
H_2 \Lambda_{D_{22}}^{-1} H_2),\quad
\Omega=\Lambda_{\rho}
\end{equation}
For given values of $Q_1$ and $Q_2$, the eigenvalues, $\nu$, of equation(\ref{Eigen1})$_1$ are obtained from
\begin{equation}\label{Eigen-2}
det\left|\Phi
-\nu^2\Omega\right|=0,
\end{equation}
and for each eigenvalue, the corresponding displacement vector $W$, is given by (\ref{Eigen1})$_1$, and the stress components by
\begin{equation}\label{T1}
T_{1}=i\Lambda_{D_{11}}^{-1}H_1 W,\qquad
T_{2}=i\Lambda_{D_{22}}^{-1}H_2 W.
\end{equation}

\subsection{Rayleigh Qutient}\label{Rayleigh}

For the Rayleigh quotient, the stress-strain relations (\ref{T1}) are written as
\begin{equation}\label{Strain}
T_{1}=i\Lambda_{\mu_{11}}H_1 W,\qquad
T_{1}=i\Lambda_{\mu_{22}}H_1 W.
\end{equation}
Substitution into the basic equilibrium equation,
$iH_1T_1+iH_2T_2+\nu^2\Lambda_{\rho} W=0$,
yields
\begin{equation}\label{Ray}
\Phi^{(Rayleigh)}=(H_1 \Lambda_{\mu_{11}}H_1+
H_2 \Lambda_{\mu_{22}} H_2).
\end{equation}
The eigenvalues are now obtained by replacing $\Phi$ in equation (\ref{Eigen-2}) by
$\Phi^{(Rayleigh)}$ and proceeding as before.

Once $W$, and $T_k$, $k=1,2$, are obtained for a frequency pass-band, the periodic parts of the field variables are as follows:
\begin{equation}\label{w}
w^p(\xi_{1},\xi_{2})=
\sum_{\alpha,\beta=-N}^{+N}W^{(\alpha \beta)}e^{i2\pi(\alpha \xi_1+\beta \xi_2)},\quad
\end{equation}
\begin{equation}\label{wdot}
\dot{w}^p(\xi_{1},\xi_{2})=-i\nu
\sum_{\alpha,\beta=-N}^{+N}W^{(\alpha \beta)}e^{i2\pi(\alpha \xi_1+\beta \xi_2)},\quad
\end{equation}
\begin{equation}\label{tau-1}
\tau_k^p(\xi_{1},\xi_{2})=
\sum_{\alpha,\beta=-N}^{+N}T_k^{(\alpha\beta)}e^{i2\pi(\alpha \xi_1+\beta \xi_2)},\quad
k=1,2,
\end{equation}

\subsection{Expressions for $\Lambda_{f(\xi_1,\xi_2)}$ in Special Cases}\label{C1}

When a rectangular unit cell contains a nested sequence of either rectangular or elliptical inclusions, matrix $\Lambda_{f(\xi_1,\xi_2)}$ can be  calculated explicitly for piecewise constant values of $f(\xi_1,\xi_2)$.  Consider an $a_1$ by $a_2$ unit cell, $\Omega_1$, that contains a nested sequence of $n-1$ concentric either elliptical or rectangular subregions,
$ \Omega_1\supset \Omega_2 \supset \Omega_3 ~  {...}  \supset \Omega_n.$

Denote the dimensions of the principal axes of a typical subregion $\Omega_j$ by ${a}_1(j)$ and ${a}_2(j)$.
Then the area of the $j^{th}$ rectangular subregion
would be $\Omega_j=a_1(j)a_2(j)$, and that of an elliptical subregion would be
$\Omega_j=\frac{\pi}{4}a_1(j)a_2(j)$.
Let $f(j)$ stand for either the mass-density or the shear modulus of the subregion $\Omega_j-\Omega_{j-1}$.  From (\ref{Operator}) now obtain,
\begin{equation}\label{Lambda}
   \Lambda_f=
	 \sum_{k=2}^{n}(f_k-f_{k-1})g_k,
\end{equation}
\begin{equation}\label{g_k1}
	g_k=\int_{\Omega_k}
	 exp\{i2\pi[(\alpha-\gamma)\xi_1+(\beta-\delta)\xi_2]\} d\xi_1d\xi_2,
\end{equation}
where for a rectangular subregion $g_k$ is given by,
\begin{displaymath}
 g_k = \left\{
  \begin{array}{lr}
 	 \frac{sin(\pi n_1a_1(k))}{\pi n_1}
	  \frac{sin(\pi n_2 a_2(k))}{\pi n_2} & n_1\ne 0,\quad n_2\ne 0,\\
        \frac{sin(\pi n_1a_1(k))}{\pi n_1}a_2(k) &  n_1\ne 0,\quad n_2 = 0,\\
        \frac{sin(\pi n_2 a_2(k))}{\pi n_2}a_1(k)&  n_1= 0,\quad n_2 \ne 0,\\
        a_1(k)a_2(k), &  n_1= 0,\quad n_2 = 0,\\
     \end{array}
   \right.
\end{displaymath}
\begin{equation}\label{g_k2}
n_1=\alpha-\gamma,\quad
n_2=\beta-\delta,\quad
(k ~ not ~ summed);
\end{equation}
and for an elliptical subregion $g_k$ becomes,
\begin{equation}
g_k=\frac{\pi}{2}\frac{a_1(k)a_2(k)J_1(R_k)}{R_k},\quad
R_k=\pi \{[n_1a_1(k)]^2+[n_2a_2(k)]^2\}^{1/2}.
\end{equation}

\section{Discussion and Conclusions}\label{conclusions}

Periodic elastic composites can be designed to have static and dynamic characteristics that are not shared by their constituent materials.  In this work we have explored  some of their uncommon acoustic properties using anti-plane shear-waves.
By superimposing the energy-flux vectors on equifrequency contours in the plane of the wave-vector components, and supplementing this with a three-dimensional graph of the corresponding frequency surface,
we have shown that
a wealth of information can be extracted essentially at a glance.
In this manner, we have revealed that a composite with even a simple square unit cell containing a central circular inclusion, when interfaced with a suitable homogeneous solid can display:
\textit{(a) negative refraction with negative phase-velocity refraction,
(b) negative refraction with positive phase-velocity refraction,
(c) positive refraction with negative phase-velocity refraction,
(d) positive refraction with positive phase-velocity refraction, or even
(e) complete reflection with no energy transmission,
depending on the frequency, direction, and the wave length of the plane-wave which is incident from the homogeneous solid to the interface.}
Hence we have shown that negative energy refraction can be accompanied by
positive phase velocity modulation (\textit{forward wave with negative refraction}).

The proposed computational tool is simple and  efficient, having a remarkable rate of convergence. To show that the solution would actually converge to the exact results, we  compared our results with those obtained using the Rayleigh quotient with very large number of plane-wave expansion series. The results of the new quotient quickly converges to the limiting values which the Rayleigh quotient needs thousands of terms to achieve.

As supplementary materials, we will provide our MatLab codes to interested readers.

\textbf{Acknowledgments}:
This research has been conducted at the Center of Excellence for Advanced Materials (CEAM) at the University of California, San Diego, under DARPA  RDECOM W91CRB-10-1-0006 to the University of California, San Diego.

\section{References}


\bibliographystyle{jphysicsB}
\bibliography{REFS_Harmonic-1}

\end{document}